\documentclass[prl,preprint,showpacs,superscriptaddress]{revtex4-1}
\usepackage{setspace}
\usepackage{amsfonts,amsmath,amssymb}
\usepackage{txfonts}
\usepackage{graphicx}
\usepackage{bm}
\usepackage{epstopdf}
\usepackage{verbatim}
\usepackage{units}
\usepackage[T1]{fontenc}
\usepackage[latin9]{inputenc}
\usepackage{amsbsy}
\usepackage{esint}

\begin{document}

\title{Statistical simulation of the magnetorotational dynamo}

\author{J.~Squire}
\affiliation{Department of Astrophysical Sciences and Princeton Plasma Physics Laboratory, Princeton University, Princeton, NJ 08543}
\author{A.~Bhattacharjee}
\affiliation{Department of Astrophysical Sciences and Princeton Plasma Physics Laboratory, Princeton University, Princeton, NJ 08543}
\affiliation{Max Planck/Princeton Center for Plasma Physics, Princeton University, Princeton, NJ 08543}

\begin{abstract}
Turbulence and dynamo induced by the magnetorotational instability (MRI) are analyzed using quasi-linear statistical 
simulation methods. It is found that homogenous turbulence is unstable to a large scale dynamo instability, which saturates to 
an inhomogenous equilibrium with a strong dependence on the magnetic Prandtl number (Pm). Despite 
its enormously reduced nonlinearity, the dependence of the angular momentum transport 
on Pm in the quasi-linear model is qualitatively similar to that of nonlinear MRI turbulence. 
This indicates that recent convergence problems may be related to
large scale dynamo and suggests how dramatically simplified models 
may be used to gain insight into the astrophysically 
relevant regimes of very low or high $\mathrm{Pm}$.
\end{abstract}



\maketitle
Understanding the complex web of nonlinear interactions that are important
for the sustenance of turbulence induced by the magnetorotational
instability (MRI) \cite{Balbus:1998tw} has proven to be a difficult
undertaking. Indeed, despite many theoretical and computational studies,
results with quantitative application to most regimes relevant
for astrophysical disks remain elusive. The basic problem is
that astrophysical objects generally contain an enormous range of
dynamically important scales, as measured by the fluid and magnetic
Reynolds numbers ($\mathrm{Re}$ and $\mathrm{Rm}$ respectively).
Of course, any simulation is necessarily
limited in its resolvable scales, and the question of whether a set
of results would change significantly with resolution becomes subtle
and very difficult to answer conclusively. In the case of MRI turbulence,
all indications are that at currently available resolutions, simulation convergence
depends on the details of the computational domain 
\cite{Fromang:2007cg,Bodo:2008fr,Shi:2010fn,Davis:2010dr,Longaretti:2010ha,Simon:2012dq,Bodo:2014fx}, and the scaling of pertinent quantities
such as the turbulent momentum transport remains unclear.
Of particular importance \cite{Fromang:2007cy,Fromang:2010es,Simon:2011bn,Oishi:2011ei} is the scaling with magnetic Prandtl number  
$\mathrm{Pm}=\mathrm{Rm}/\mathrm{Re}$;
astrophysical objects invariably have very high or low $\mathrm{Pm}$
but these regimes are extremely computationally challenging. Indeed, it is currently unclear whether MRI turbulence at very low Pm
is sufficiently virulent to explain the accretion rate inferred from
luminosity observations of compact objects, since
turbulent activity seems to decrease with $\mathrm{Pm}$ or disappear
altogether \cite{Fromang:2007cy,Longaretti:2010ha,Flock:2012dn}
(but see Refs.~\cite{Kapyla:2011fj,Oishi:2011ei}). A large-scale dynamo generating strong
azimuthal magnetic fields \cite{Brandenburg:1995dc,Hawley:1996gh,Lesur:2008cv,Gressel:2010dj,Simon:2012dq,Ebrahimi:2014jt}
seems to be a key component of the turbulence,
although the exact nature of the interactions and importance of different
effects (e.g., vertical stratification, compressibility) remains
unclear. 

In this letter we study MRI turbulence and dynamo in the zero net-flux
unstratified shearing box using novel quasi-linear statistical simulation
methods
(from hereon we shall use the term second-order cumulant expansion
(CE2) \cite{Marston:2008gx}, although the term stochastic structural stability
theory (S3T) \cite{Farrell:2003ud} is also common and pertains to similar ideas).
This involves driving
an ensemble of linear fluctuations in mean fields that depend only
on the vertical co-ordinate ($z$), with the nonlinear
stresses of these fluctuations self-consistently driving evolution
of the mean fields. Our motivation for this is two-fold: Firstly,
despite being a rather recent subject, direct statistical simulation
-- the method of simulating flow \emph{statistics} rather than an
individual realization -- has proven to be a useful computational
technique in 
a variety of applications \cite{Farrell:2012jm,Tobias:2011cn,Srinivasan:2012im,Parker:2013hy}. 
An equilibrium of the system is in general a \emph{turbulent} state, and analysis
of its stability properties and bifurcations can be very rewarding.  
Secondly, fully developed MRI turbulence is incredibly complex and
we feel there is much useful insight to be gained by selectively \emph{removing}
important physical effects in the hope of discovering underlying
principles. Motivated by the idea that strong linear MRI growth is 
possible at all scales due to nonmodal effects \cite{Squire:2014cz},
our quasi-linear model involves neglecting almost
all of the nonlinear interactions in the system and can easily be
systematically reduced further.  

Remarkably, despite the strongly reduced nonlinearity, we demonstrate that the qualitative dependence
of saturated CE2 states on $\mathrm{Pm}$ is the same as
nonlinear MRI turbulence. In particular, at fixed magnetic Reynolds
number ($\mathrm{Rm}$), an increase in $\mathrm{Pm}$ causes an increase
in the intensity of the turbulence (as measured by the angular momentum
transport), despite the fact that the system is becoming more
dissipative. This illustrates that the strong $\mathrm{Pm}$ dependence
of the MRI \cite{Fromang:2007cy} is (at least partially) due to increased large-scale
dynamo action at higher $\mathrm{Pm}$; this is the only physical
effect retained in the CE2 model beyond simple excitation
of linear waves (which show the opposite trend). As discussed, CE2 is very well suited to the study of 
bifurcations between turbulent states of the system. We see two important 
bifurcations -- the first marking the onset of a dynamo instability of homogenous turbulence, the second
a transition to a time-dependent state -- and the $\mathrm{Pm}$ dependence of 
several aspects of these transitions is strongly suggestive. It is our hope that
gaining insight into the cause of such behavior will allow extrapolation to the most 
astrophysically relevant low/high $\mathrm{Pm}$ regimes.
Note that the approach is quite distinct from, and complementary to, previous
nonlinear dynamics work on MRI dynamo \cite{Rincon:2007bm,Riols:2013dk},
which has focused on searching for cycles in the
full nonlinear system at low Rm. Strong similarities can
be drawn between the dynamo mechanisms identified in these works
and magnetic field generation in our CE2 model \cite{Farrell:2012jm}.

The starting point of our study is the local incompressible MHD equations
in a shearing background in the rotating frame,
\begin{align}
\frac{\partial\bm{u}}{\partial t} & -q\Omega x\frac{\partial\bm{u}}{\partial y}+\left(\bm{u}\cdot\nabla\right)\bm{u}+2\Omega\bm{\hat{z}}\times\bm{u}=\nonumber \\
 & -\nabla p+\bm{B}\cdot \nabla \bm{B}+q\Omega u_{x}\bm{\hat{y}}+\bar{\nu}\nabla^{2}\bm{u},\nonumber \\
\frac{\partial\bm{B}}{\partial t} & -q\Omega x\frac{\partial\bm{B}}{\partial y}=-q\Omega B_{x}\bm{\hat{y}}+\nabla\times\left(\bm{u}\times\bm{B}\right)+\bar{\eta}\nabla^{2}\bm{B},\nonumber \\
 & \nabla\cdot\bm{u}=0,\;\;\;\nabla\cdot\bm{B}=0.\label{eq:NL MHD equations}
\end{align}
These are obtained from the standard MHD equations for a disk with
radial stratification by considering a small Cartesian volume (at
$r_{0}$) co-rotating with the fluid at angular velocity $\Omega\left(r\right)\sim\Omega_{0}r^{-q}$. 
In this limit the velocity shear is linear,
$\bm{U}_{0}=-q\Omega x\bm{\hat{y}}$, and $\bm{u}$ denotes velocity
fluctuations about this background. The directions $x,\, y,\, z$
in Eq.~\eqref{eq:NL MHD equations} correspond respectively to the
radial, azimuthal and vertical directions in the disk. The use of dimensionless variables in Eq.~\eqref{eq:NL MHD equations}
means $\Omega\equiv\Omega\left(r_{0}\right)=1$, and the bulk flow Reynolds numbers
 are $\mathrm{Re}=q/\bar{\nu}$,
$\mathrm{Rm}=q/\bar{\eta}$. Throughout
this work we consider a homogenous background (no vertical stratification),
with zero net magnetic flux, and use shearing box boundary conditions
(periodic in $y,\,z$, periodic in the shearing frame in $x$)
with an aspect ratio $\left(L_{x},L_{y},L_{z}\right)=\left(1,\pi,1\right)$.

The basis of our application of CE2 to MRI turbulence is
a splitting of Eq.~\eqref{eq:NL MHD equations} into its mean and
fluctuating parts, as defined by the horizontal average, 
$\overline{f\left(\bm{x}\right)}\left(z\right)\equiv (L_{x}L_{y})^{-1}\int dxdy\, f\left(\bm{x}\right)$.
This averaging is chosen because it is the simplest possible that
allows for the strong $z$-dependent $B_y$ observed
in nonlinear simulations \cite{Lesur:2008cv,Kapyla:2011fj}.
Schematically representing the state of the system $\left(\bm{u},\bm{B},P\right)$
as $U$, a decomposition of Eq.~\eqref{eq:NL MHD equations} into
equations for $\bar{U}$ and $u'=U-\bar{U}$ gives\begin{subequations}
\begin{gather}
\partial_{t}\bar{U}=\mathcal{A}_{mean}\cdot\bar{U}+\overline{\mathcal{R}\left(u',u'\right)},\label{eq:QL mean}\\
\partial_{t}u'=\mathcal{A}_{fluct}\left(\bar{U}\right)\cdot u'+\left\{ \mathcal{R}\left(u',u'\right)-\overline{\mathcal{R}\left(u',u'\right)}\right\} +\xi_{t}.\label{eq:QL fluct}
\end{gather}
\end{subequations}Here $\mathcal{A}_{mean}$ and $\mathcal{A}_{fluct}\left(\bar{U}\right)$
are the linear operators for the mean and fluctuating parts, $\mathcal{R}\left(u',u'\right)$
represents the nonlinear stresses, and $\xi_{t}$
is an additional white-in-time driving noise term. The principle approximation, 
which is key to CE2 and leads to the \emph{quasi-linear}
system, is to neglect the eddy-eddy nonlinearity $\left\{ \mathcal{R}\left(u',u'\right)-\overline{\mathcal{R}\left(u',u'\right)}\right\} $
in Eq.~\eqref{eq:QL fluct}, causing the only nonlinearity
to arise from the coupling to Eq.~\eqref{eq:QL mean}. The driving
noise $\xi_{t}$ can be considered either a physical source of noise \cite{Marston:2008gx},
or a particularly simple closure representing the effects of the neglected nonlinearity \cite{Farrell:2003ud}.

Rather than evolving the non-deterministic Eq.~\eqref{eq:QL fluct}, 
consider the single time correlation matrix of an ensemble
of fluctuations
$\mathcal{C}_{ij}\left(\bm{x}_{1},\bm{x}_{2},t\right)=\left\langle u'_{i}\left(\bm{x}_{1},t\right)u'_{j}\left(\bm{x}_{2},t\right)\right\rangle$,
where $\left\langle \cdot\right\rangle $ denotes the average
over realizations of $\xi_t$. Multiplying Eq.~\eqref{eq:QL fluct}
by $\partial_{t}u\left(\bm{x}_{2}\right)$ followed by an ensemble
average leads to \cite{Farrell:2003ud,Srinivasan:2012im}
\begin{equation}
\partial_{t}\mathcal{C}=\mathcal{A}_{fluct}\left(\bar{U}\right)\cdot\mathcal{C}+\mathcal{C}\cdot\mathcal{A}_{fluct}\left(\bar{U}\right)^{\dagger}+\mathcal{Q},\label{eq:C eqn}
\end{equation}
where $\mathcal{Q}=\left\langle \xi\left(\bm{x}_{1},t\right)\xi\left(\bm{x}_{2},t\right)\right\rangle $
is the spatial correlation of the noise 
\footnote{Eq.~\eqref{eq:C eqn}, the basis for the CE2 system, can also
be derived as a truncation of the cumulant expansion (in an inhomogenous background) at second order \cite{Tobias:2011cn}. 
This can be generalized to yield higher order statistical equations that are often similar to inhomogenous 
versions of well known closure models, e.g., the eddy-damped quasi-normal Markovian
approximation \cite{Marston:2012ar}.}.
Using homogeneity in  $x,\,y$, Eq.~\eqref{eq:C eqn} can be reduced
to 4 dimensions with the change of variables, $x=x_{1}-x_{2}$, $y=y_{1}-y_{2}$.
Assuming ergodicity -- the equivalence of the $x,y$ and ensemble
averages -- the nonlinear stresses $\overline{\mathcal{R}\left(u',u'\right)}$
in the mean field equations {[}Eq.~\eqref{eq:QL mean}{]} can be
calculated directly from $\mathcal{C}$. With this
change Eqs.~\eqref{eq:QL mean} and \eqref{eq:C eqn} comprise the
CE2 system. Aside from the noise, conservation laws are inherited from the nonlinear
system (e.g., energy, magnetic helicity).

The MRI mean field equations are very simple,
\begin{gather}
\partial_{t}\left(\bar{U}_{x},\bar{U}_{y}\right)=\left(2\bar{U}_{y},\left(q-2\right)\bar{U}_{x}\right)+\left(\mathcal{R}_{x},\mathcal{R}_{y}\right)\nonumber \\
\partial_{t}\left(\bar{B}_{x},\bar{B}_{y}\right)=\left(0,-q\bar{B}_{x}\right)+\left(\mathcal{M}_{x},\mathcal{M}_{y}\right),\label{eq:MRI CE2 eqns}
\end{gather}
with $\partial_z \bar{U}_{z}=\partial_z \bar{B}_{z}=0$ due to the divergence constraints.
The nonlinear stresses arising from the fluctuating variables, $\mathcal{R}_{j}=\overline{\left\langle -\left(\bm{u}'\cdot\nabla\bm{u}'\right)_{j}+\left(\bm{b}'\cdot\nabla\bm{b}'\right)_{j}\right\rangle }$
and $\mathcal{M}_{j}=\overline{\left\langle \left(\nabla\times\left(\bm{u}'\times\bm{b}'\right)\right)_{j}\right\rangle },$
are calculated by applying appropriate derivative operators
to the $\mathcal{C}$ matrix. We solve for $\mathcal{C}$ in the variables, $u\equiv u_{x}',$ $b\equiv b_{x}',$
$\zeta\equiv\partial_{z}u_{y}'-\partial_{y}u_{z}',$ $\eta\equiv\partial_{z}b_{y}'-\partial_{y}b_{z}'$,
which conveniently reduces the dimension of $\mathcal{C}$
and removes divergence constraints. The equations, however, become very complex and we do not reproduce
them here (\emph{Mathematica} scripts are used to automatically generate the 
required C++ code
\footnote{\emph{Mathematica} scripts, along with the
automatically generated CE2 equations and C++ code, can be
found in the `Mathematica' folder of https://github.com/jonosquire/MRIDSS}). 
We use a Fourier pseudo-spectral method (with $3/2$ dealiasing)
in the shearing frame with the remapping method of Ref.~\cite{Lithwick:2007ge}, 
and a semi-implicit Runge-Kutta time-integrator. 

In all calculations presented here, we initialize with $\mathcal{C}=0$.
The spatial correlation of $\xi_t$ is chosen to drive each
mode equally in energy \cite{Farrell:2012jm}, multiplied
by an amplitude factor $f_{\xi}$. While we have explored the dependence on $f_{\xi}$, 
for simplicity all calculations
in this letter use the same value ($f_{\xi}=4$ in our normalization)
and we change the physical parameters $\mathrm{Rm}$ and $\mathrm{Pm}$
to illustrate bifurcations of the system. For reference, this noise
level drives homogenous turbulence at $\mathrm{Rm}=12000,$ $\mathrm{Pm}=1$
to a mean total energy of $\sim\!\!\!0.05$. $\mathrm{Rm}=12000$ computations use the resolution $40\times80\times(4\times64)^{2}$ 
(note that dealiasing is not required in $x$ and $y$). To ensure 
accuracy we have tested conservation of energy, as well as doubling the resolution (to $80\times160\times(4\times128)^{2}$) for $\mathrm{Pm}=1,\, 4$.

\paragraph{The MRI dynamo instability}

In contrast to the original MRI equations {[}Eq.~\eqref{eq:NL MHD equations}{]},
a general stable equilibrium of the CE2 system {[}Eqs.~\eqref{eq:C eqn}
and \eqref{eq:MRI CE2 eqns}{]} corresponds to a statistically stationary
turbulent state within the quasi-linear approximation. If such
an equilibrium is rendered unstable by a change in system parameters,
this turbulent state is no longer possible and a rearrangement of
the mean fields and flow statistics will occur. This type of instability
has no counterpart in standard MHD stability theory; it pertains to
the idea that the collective effect of the ensemble of fluctuating
fields is to re-enforce perturbations to the mean fields through the
nonlinear stresses, causing instability. Of course, such ideas are familiar
in mean-field electrodynamics \cite{1978mfge.book.....M}, and the CE2 method seems  
well suited for more general study of large scale dynamos.  

Homogenous turbulence, with $\left(\bar{\bm{U}},\bar{\bm{B}}\right)=0$,
is the simplest non-trivial equilibrium of the CE2 MRI system, with
all nonlinear stresses
vanishing identically. However, at fixed noise, as Re and Rm are increased
from zero this equilibrium becomes unstable around $\mathrm{Rm}\approx 1500$ (this value changes 
with noise level). 
\begin{figure}
\centering{}\includegraphics[width=0.7\columnwidth]{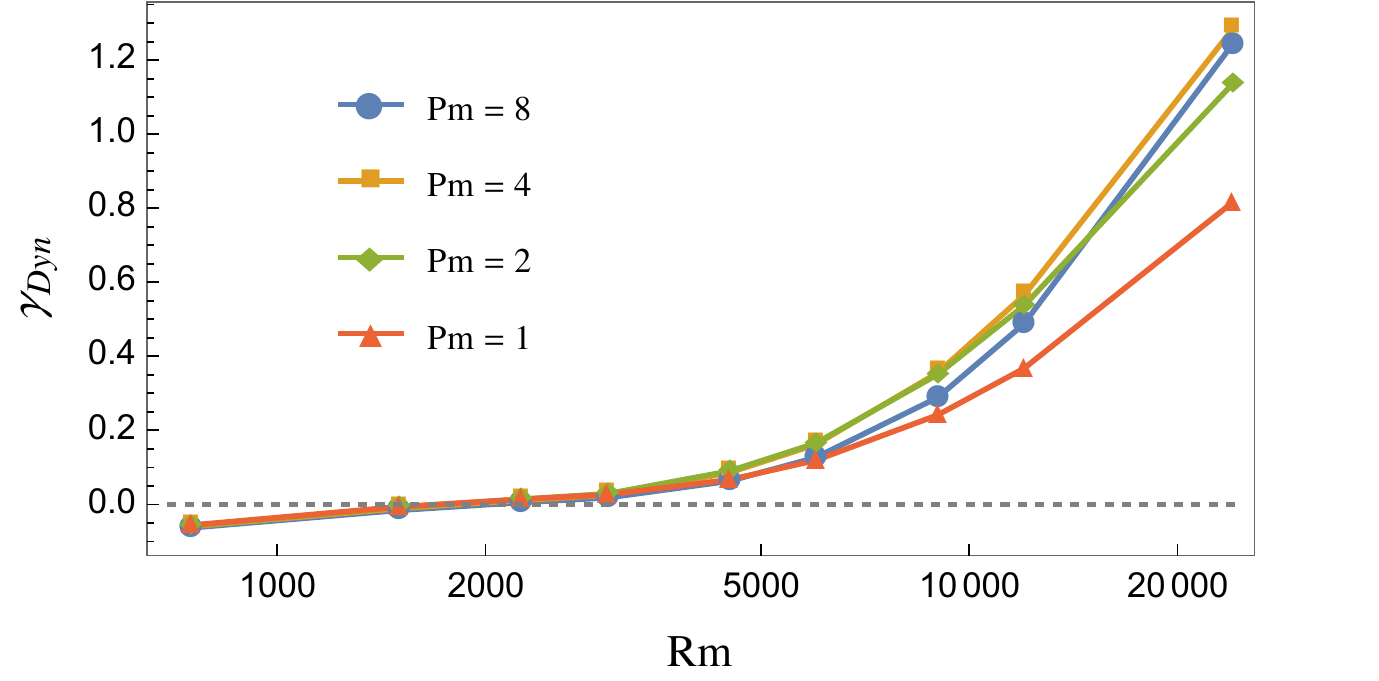}
\caption{Growth rate $\gamma_{Dyn}$ of the mean field, $\bar{B}_{y}=\bar{B}_{y0}\left(z\right)e^{\gamma_{Dyn}t}$,
as a function of magnetic Reynolds number at $\mathrm{Pm}=1,\,2,\,4\,\mathrm{and}\,8$. \label{fig:Linear Growth}}
\end{figure}
Such behavior is illustrated in Fig.~\ref{fig:Linear Growth}, which
shows the growth rate $\gamma_{Dyn}$ of this dynamo instability. This is calculated by first evolving
Eq.~\ref{eq:C eqn} to the homogenous equilibrium by artificially
removing the nonlinear feedback, then introducing a very small
($\sim\!\!10^{-15}$) random mean field (with the amplitudes of $\bar{\bm{U}},\,\bar{B}_{x}$
1/10 that of $\bar{B_{y}}$). (While it is possible to solve for the Floquet eigenspectrum
directly, this is challenging due to the grid size.)
Following the introduction of mean-field feedback there is
a sustained period of exponential growth in $\bar{\bm{B}}$ for $\mathrm{Rm}\gtrsim1500$. The observed
eigenmodes are sinusoidal in $z$ (ensured by spatial homogeneity)
although not generally the largest mode in the box, satisfy $B_{x}\ll B_{y}$
and seem to have $\bar{\bm{U}}=0$ %
\footnote{This may not be the case at the highest $\mathrm{Rm}$ studied, since
$\bar{\bm{U}}$ grows slowly but does not ever get small enough relative
to $\bar{\bm{B}}$ to say for sure whether the eigenmode satisfies
$\bar{\bm{U}}=0$ or just $\bar{\bm{U}}\ll\bar{\bm{B}}$. In either
case, it is far too small to be of dynamical importance in the linear
growth phase. %
}. While it is certainly expected that $\gamma_{Dyn}$
increase strongly with $\mathrm{Rm}$ -- fluctuations grow to
a higher amplitude and there is less $\bar{\bm{B}}$ dissipation -- its dependence on $\mathrm{Pm}$
is more interesting and suggestive. An increase in $\mathrm{Pm}$ implies
more dissipation (through increasing
$\bar{\nu}$), yet Fig.~\ref{fig:Linear Growth} shows that $\gamma_{Dyn}$
can increase, particularly at higher $\mathrm{Rm}$. In addition, 
$\tfrac{\partial}{\partial \mathrm{Rm}}\gamma_{Dyn}\left(\mathrm{Rm}\right)$ increases with
$\mathrm{Pm}$, with potentially interesting consequences for the
very high $\mathrm{Rm}$ limit. The instability is driven by the radial 
stress $\mathcal{M}_{x}$ causing an increase in $\bar{B}_{x}$, which
in turn drives $\bar{B}_{y}$ through the $\Omega$ effect, $-q\bar{B}_{y}$
{[}see Eq.~\eqref{eq:MRI CE2 eqns}{]}. The effect of the azimuthal
stress $\mathcal{M}_{y}$ is always negative. This is identical to
the dynamo mechanism studied in detail in Refs.~\cite{Lesur:2008cv,Lesur:2008fn},
and has strong similarities to exact nonlinear dynamo solutions at low $\mathrm{Rm}$ \cite{Rincon:2007bm,Riols:2013dk}.

\begin{figure}
\begin{centering}
\includegraphics[width=0.7\columnwidth]{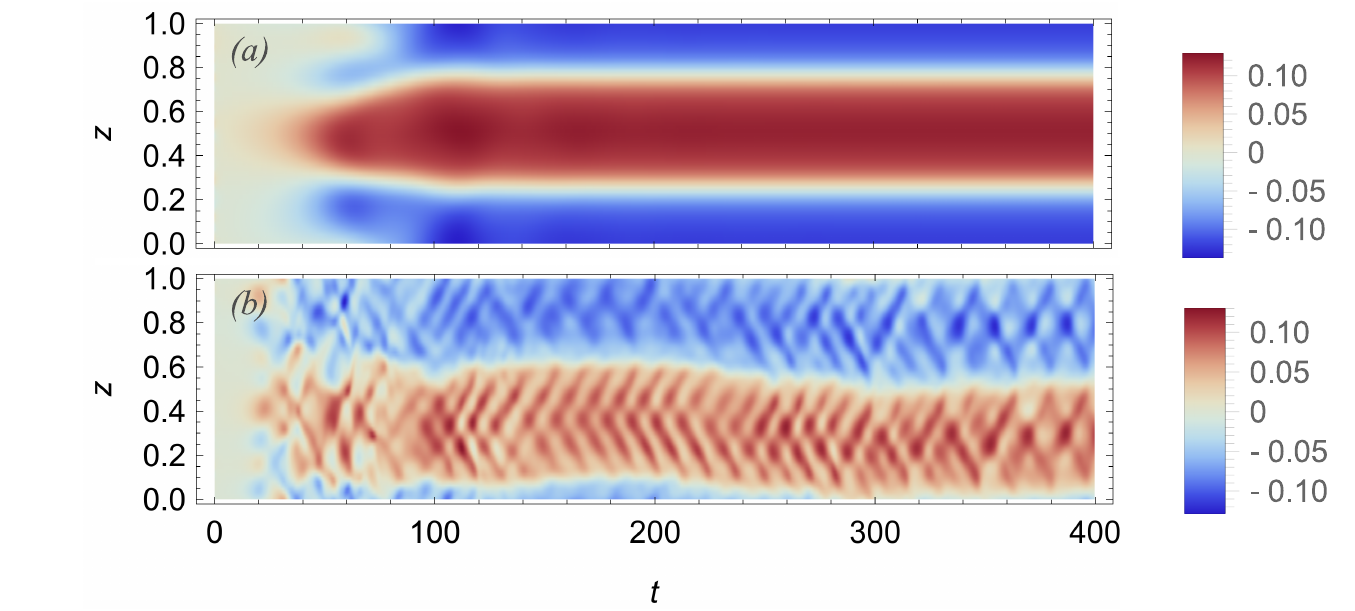}
\par\end{centering}

\begin{centering}
\includegraphics[width=0.7\columnwidth]{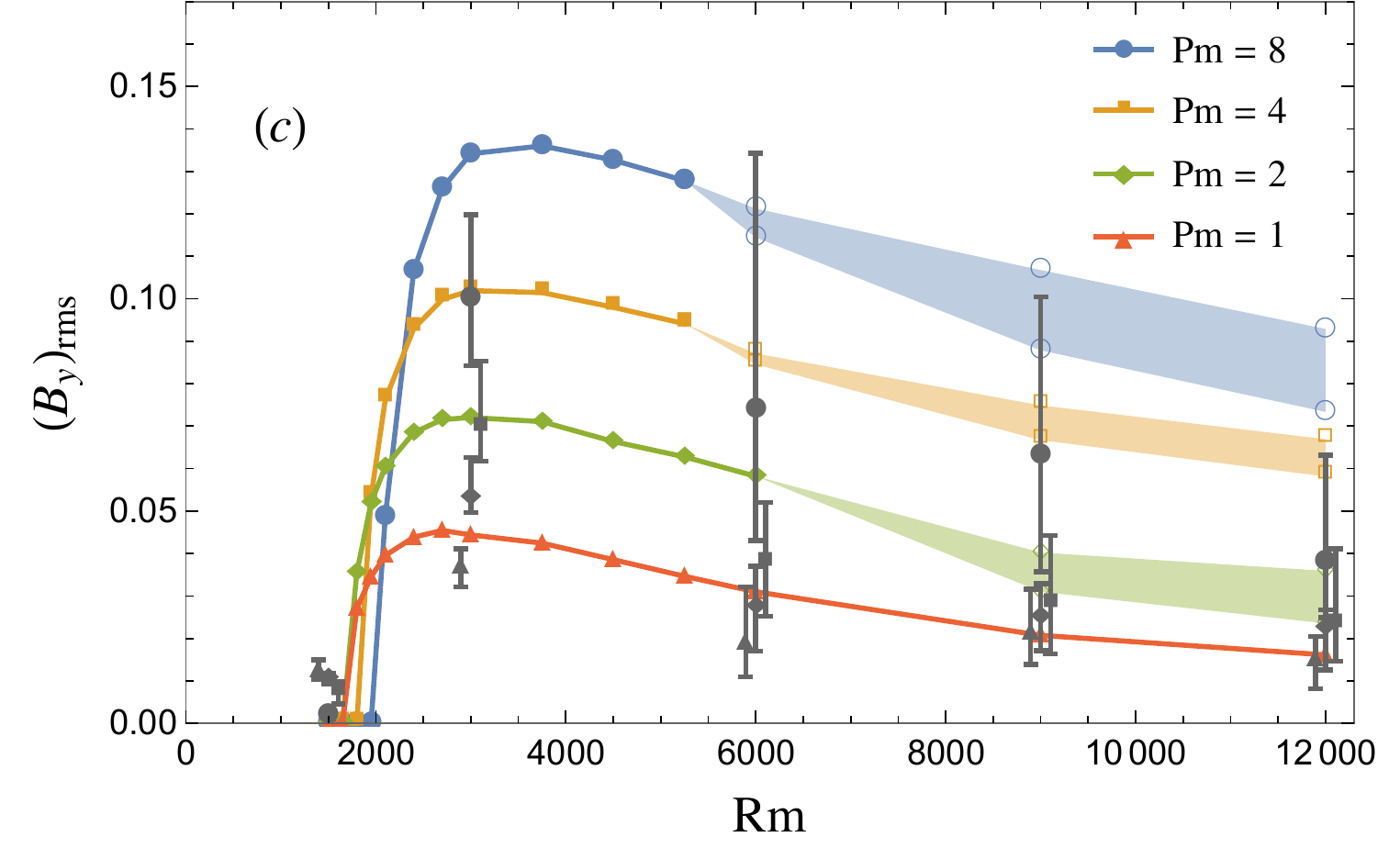}
\par\end{centering}

\centering{}\caption{Evolution of $\bar{B}_y$ as a function of $\left(z,t\right)$ at $\mathrm{Pm}=4$
for (a) $\mathrm{Rm}=4500$, time-independent saturated state, and
(b) $\mathrm{Rm}=12000$, time-dependent saturated state. (c) Magnitude
of $\bar{B}_{y}$ as measured by $\left( \bar{B}_y\right)_{\mathrm{rms}} = \left( \tfrac{1}{L_{z}}\int dz\,\left|\bar{B}_{y}\right|^{2} \right)^{1/2}$ at
saturation, as a function of $\mathrm{Rm}$ and $\mathrm{Pm}$.
The shaded regions illustrate the approximate maxima and minima of
the time-dependent $\bar{B}_{y}$ when the system did not reach a
time-independent statistical equilibrium. 
Gray points (point styles as for CE2 results) illustrate the mean values of equivalent driven nonlinear simulations,
with error-bars illustrating the approximate maxima and minima (the slight horizontal offset of $\mathrm{Pm}=1,\,4$ points
is for clarity, the same Rm is used for all Pm).\label{fig:By sat}}
\end{figure}
Of more relevance to fully developed turbulence are the saturation
characteristics of the dynamo instability. To save computation, we initialize with moderately
strong random mean fields (amplitude of $\bar{B}_{y}\approx0.01,$
$\bar{B}_{x}$ and $\bar{\bm{U}}$ initialized at 1/10 that of $\bar{B}_{y}$ --
we have also studied initialization with the largest mode of the
box obtaining similar results). As $\mathrm{Rm}$ is increased and 
homogenous equilibrium rendered unstable, the system saturates to a
new CE2 equilibrium with a strong background $\bar{B}_{y}$  that varies on the largest
scale in the box, as illustrated by the  example
in Fig.~\ref{fig:By sat}(a). As we increase $\mathrm{Rm}$ further,
a second bifurcation occurs, at which the inhomogenous equilibrium
appears to become unstable and the system transitions to a quasi-periodic
time-dependent state. An example of this state, which occurs more
readily at higher $\mathrm{Pm}$, is shown in Fig.~\ref{fig:By sat}(b).
These two bifurcations -- first to an inhomogenous state dominated
by mean fields, then the loss of equilibrium of this state -- bear
a strong resemblance to the transitions seen in hydrodynamic plane
Couette flow \cite{Farrell:2012jm}, in which the second transition
is associated with self-sustaining behavior. Such a self-sustaining
process is not possible within our model due to 
the choice of 1-D mean-fields (as opposed to 2-D in Ref.~\cite{Farrell:2012jm}), but the similarity
as well as its $\mathrm{Pm}$ dependence is striking. Understanding
physical mechanisms behind the loss of equilibrium may
give useful insights into the self-sustaining dynamo that is so fundamental
to zero net-flux turbulence. 

This information is presented more compactly in Fig.~\ref{fig:By sat}(c),
which illustrates the saturated $\bar{B}_{y}$ amplitude over a range of $\mathrm{Rm},\,\mathrm{Pm}$. The dependence of
the saturated amplitude on $\mathrm{Pm}$ is enormous (contrary to
previous results on the large scale dynamo \cite{Brandenburg:2009ii}), and
can be well understood at low $\mathrm{Rm}$ using the linear properties of inhomogenous
shearing waves \cite{Lesur:2008cv,Lesur:2008fn}. Also shown is the
mean azimuthal field $\overline{B_y \left(\bm{x}\right)}\left(z\right)\equiv (L_{x}L_{y})^{-1}\int dxdy\, B_y\left(\bm{x}\right)$
in driven nonlinear simulations (using statistically equivalent noise to that in CE2),
which shows the same trends although 
amplitudes are somewhat smaller.  These simulations are run
at a resolution $64\times 128 \times 64$ ($\mathrm{Rm}\leq 9000$) and $128\times 256 \times 128$ ($\mathrm{Rm}=12000$) using the 
SNOOPY code \cite{Lesur:2007bh}, and
mean values are obtained through time averages from $t=200\rightarrow 400$. The large error-bars 
on these results illustrate how statistical simulation can be very profitable for  observing such trends in data.
Note that in contrast to most nonlinear simulation, the
driving noise extends to the smallest scales available. Future work will 
explore how the turbulent dynamo changes as this is altered in both CE2 and nonlinear simulation \cite{Constantinou:2014kb}. 
Interestingly, there is a marked
\emph{decrease }in the saturated amplitudes at all $\mathrm{Pm}$
as $\mathrm{Rm}$ is increased. We have been unable to find a convincing
physical mechanism to explain this effect, but note that it depends
critically on the interaction of the fluctuating fields with $\bar{B}_{x}$. This illustrates that some 
important physical effects may be absent from the saturation mechanism proposed in Refs.~\cite{Lesur:2008cv,Lesur:2008fn}. 

\begin{figure}
\centering{}\includegraphics[width=0.7\columnwidth]{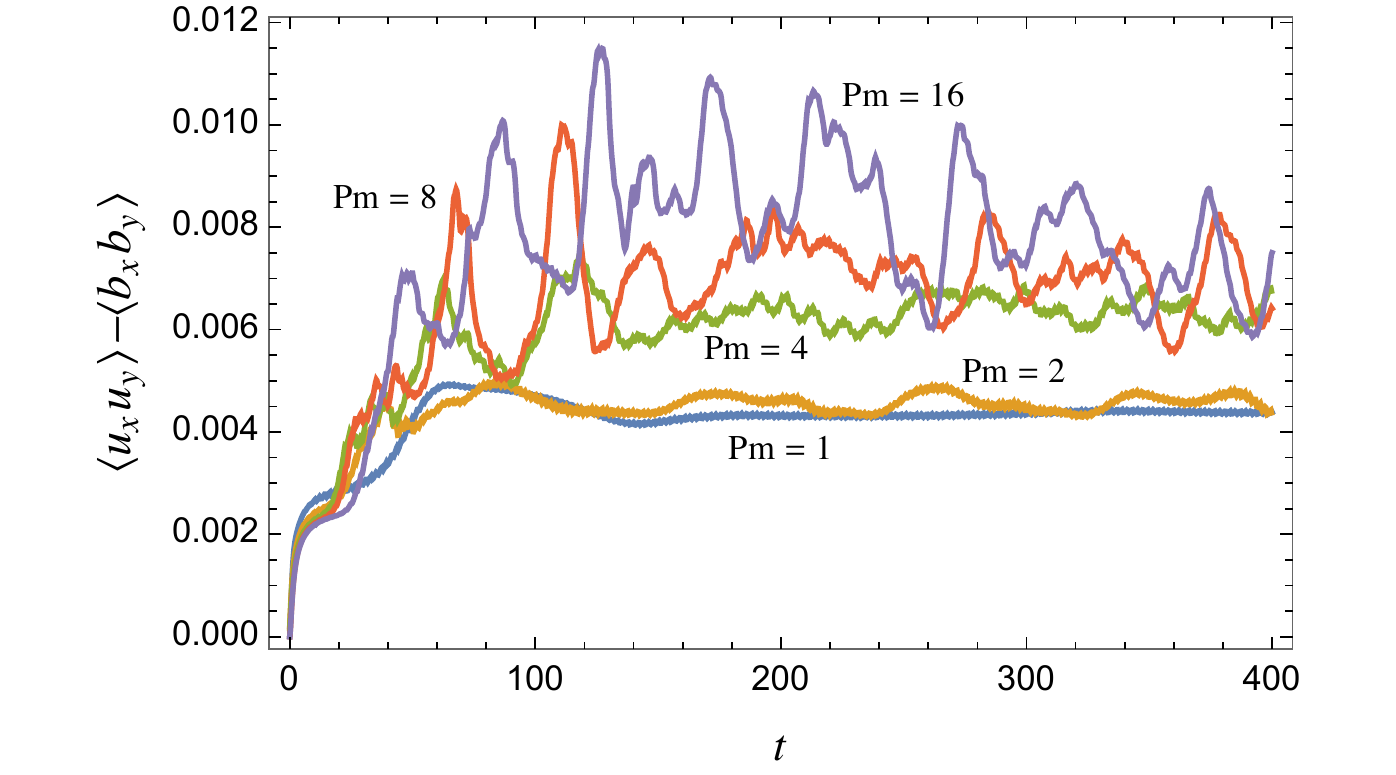}
\caption{Angular momentum transport $\left\langle u_{x}u_{y}\right\rangle -\left\langle b_{x}b_{y}\right\rangle $
(including mean and fluctuating variables) as a function of time for
$\mathrm{Rm}=12000$, $\mathrm{Pm}=1\rightarrow16$.\label{fig:Angular-momentum-transport}}
\end{figure}
In Fig.~\ref{fig:Angular-momentum-transport} we present the angular
momentum transport as a function of time for the highest $\mathrm{Rm}$ calculations presented
in Fig.~\ref{fig:By sat}. The increase in transport with $\mathrm{Pm}$
despite the increased dissipation is evident, suggesting
a relationship between shearing box convergence problems \cite{Fromang:2007cy,Fromang:2010es}
and the large scale dynamo. While the scaling is not so pronounced
as self-sustained non-linear turbulence (see e.g., Ref.~\cite{Fromang:2007cy}
figure 7), this is to be expected since the CE2 calculations are driven. 
The scaling in our driven nonlinear simulations (see Fig.~\ref{fig:By sat}, not shown in Fig.~\ref{fig:Angular-momentum-transport})
is similar, although the overall transport level is a factor of $\sim\!\!1.5$ smaller.
Note that the increase in transport is
not primarily from the mean fields directly (e.g., through $\left<\bar{B}_x \bar{B}_y \right>$),
but rather due to the fluctuations becoming more intense
as a consequence of the stronger mean fields.

\paragraph*{Discussion}

Our primary motivation for this work has been to disentangle the important
processes involved in MRI turbulence and dynamo. With this aim, we
have enormously reduced the nonlinearity of the unstratified shearing
box system, keeping only those interactions that involve
the $k_{x}=k_{y}=0$ modes (the mean fields). This removes the usual
turbulent cascade, although fluctuations are still swept to the 
smallest scales by the mean shear. Our primary result is that despite this huge simplification
-- the only nonlinearity is due to the mean field dynamo -- the CE2
system displays qualitatively similar trends to fully developed MRI turbulence.
In particular, a decrease in $\mathrm{Re}$ at fixed $\mathrm{Rm}$
($i.e.,$ an increase in $\mathrm{Pm}$), causes an increase in angular
momentum transport. This work illustrates the relationship of this trend to the large scale dynamo 
and facilitates
future analytic studies to understand the primary causes for such behavior.
The hope is that such understanding would allow extrapolation
into the high and low Pm regimes that are so computationally challenging.
In addition, statistical simulation (i.e., CE2) \cite{Farrell:2003ud,Marston:2008gx}
provides very clear information on the bifurcations between turbulent
states of the system. We see two important bifurcations as $\mathrm{Rm}$
is increased: the first is the transition from stable homogenous turbulence
to a stable inhomogenous equilibrium with strong mean-fields (the
dynamo instability), the second a loss of stability of the inhomogenous
equilibrium and transition to a near-periodic time-dependent state.
Given the strong dependence of both the saturated states and the second
bifurcation on $\mathrm{Pm}$, as well as the marked similarity to
studies of plane Couette flow \cite{Farrell:2012jm},
it seems likely that further study of this dynamo instability will
yield important insights into the fundamental nature of the MRI system.


\begin{acknowledgments}
We extend thanks to Jim Stone, Jiming Shi and John Krommes for enlightening discussion. 
This work was supported by Max Planck/Princeton Center for Plasma Physics and  U.S. DOE (DE-AC02-09CH11466).
\end{acknowledgments}

%

\end{document}